\documentclass[smallextended]{svjour3}
\pdfoutput=1  
\smartqed  
\usepackage{graphicx}
\usepackage{latexsym}
\usepackage{bbm,epsf,graphicx,epsfig,psfrag,bm}
\usepackage{mathrsfs,amsmath,amssymb,textcomp}
\usepackage{color}
 \journalname{Optical and Quantum Electronics}
\begin{document}

\title{Effective microscopic theory of quantum dot superlattice solar cells}

\author{U. Aeberhard}


\institute{U. Aeberhard \at
              IEK-5 Photovoltaik \\
              Forschungszentrum J\"ulich\\
              D-52425 J\"ulich, Germany\\
              Tel.: +49 2461 61-2615\\
              Fax: +49 2461 61-3735\\
              \email{u.aeberhard@fz-juelich.de}           
}

\date{Received: date / Accepted: date}

\maketitle

\begin{abstract} 
We introduce a quantum dot orbital tight-binding non-equilibrium Green's function approach for the
simulation of novel solar cell devices where both absorption and conduction are mediated by quantum
dot states. By the use of basis states localized on the quantum dots, the computational real space
mesh of the Green's function is coarse-grained from atomic resolution to the quantum dot spacing,
which enables the simulation of extended devices consisting of many quantum dot layers.
\end{abstract}

\section{Introduction}
Extended quantum dot superlattices (QDSL) as found in third generation solar cell devices 
\cite{green:00,marti:06} consist of a large number of weakly coupled quantum dots, with a number of
involved atoms exceeding the limit of what may still be handled by even today's largest
supercomputers. For that reason, atomistic \cite{kirchartz:09_SL} and even microscopic
continuum \cite{jiang:06} simulations of QDSL absorbers make use of the symmetry and periodicity
properties present in the idealized structure. However, in real devices, the finite number of periods, large built-in
fields and any kind of disorder make the electronic structure deviate considerably from the minibands found for
the latter and hence disqualify such simplifications. Furthermore, even for narrow electronic minibands
and gaps in the phonon spectrum, the charge carriers will be subject to various kinds of scattering, causing dissipation and
preventing purely coherent transport in the device. For the determination of the device characteristics it is thus desirable to derive a computational
model that on the one hand can still be handled numerically by making use of the localized nature of the QD wave
functions, and which on the other hand is able to cope with all of the above mentioned deviations from the idealized
structure, thanks to being based on a quantum-kinetic theory of photogeneration,  
recombination and transport, including injection and extraction of carriers at contacts,  similar to  the
non-equilibrium Green's function (NEGF) approach developed for quantum well solar cells in
\cite{ae:prb_08,ae:nrl_11,ae:jcel_review}. The present paper introduces such a model and applies it to the
investigation of (photo)carrier localization and extraction in selectively contacted silicon quantum dot absorber arrays
under varying internal field conditions.

\section{Theoretical approach}   
The NEGF theory that forms the core of the model provides a quantum version of the Boltzmann
transport equation for the charge carriers \cite{kadanoff:62,keldysh:65,datta:95,binder:95}, which includes tunneling
transport and non-local rate terms 	for scattering, generation and recombination, which are all essential processes in
the photovoltaic 	regime of device operation. For the representation of the electronic system, the present hybridization
approach 	uses the perturbative expansion of the QDSL wave function in terms of the eigenstates of the isolated dots.
The resulting molecular orbital approach can be interpreted as a tight-binding (TB) theory with quantum dot orbitals
replacing the atomic orbitals, and is in spirit similar to the first NEGF models of quantum well superlattices for quantum cascade laser simulations
\cite{wacker:02,lee:prb_02} and to NEGF models of polaron transport in QD superlattices
\cite{vukmirovic:07}. The orbitals $m=1,2,\ldots,M$ of QD $i=1,2,\ldots,N_{D}$ with energies  $\varepsilon_{im}$ are
used to represent the field operators for single particle quantum dot states, 
$\hat{\Psi}(\mathbf{r},t)=\sum_{i,n}\psi_{in}(\mathbf{r})\hat{d}_{in}(t)$, where $\hat{d}_{in}$
annihilates an electron in state $|n\rangle$ on the QD at position $\mathbf{R}_{i}$. The
non-interacting nearest-neighbor tight-binding Hamiltonian in this QD orbital basis is
\begin{align}
\mathcal{H}_{0}(t)=&\sum_{\langle i,j\rangle}\sum_{m,n=1}^{M}
t_{im,jn}\hat{d}_{im}^{\dagger}(t)\hat{d}_{jn}(t)+\sum_{i=1}^{N_{D}}\sum_{m=1}^{M}
\tilde{\varepsilon}_{im}\hat{n}_{im}(t),\label{eq:ham}
\end{align}
where $\langle i,j\rangle$ are nearest-neighbor sites, $\mathbf{t}$ is the
hopping matrix, $\hat{n}$ the density operator and
$\tilde{\varepsilon}_{im}=\varepsilon_{im}+\bar{U}_{i}$, with $\bar{U}_{i}$ the average value for the Hartree potential
of Coulomb interaction at the dot position. The heterostructure potential does no longer appear explicitly, since it has been considered in the determination of the
TB-parameters $\boldsymbol{\varepsilon}$ and $\mathbf{t}$,
\begin{align}
t_{im,jn}=&-\int d^{3}r\psi_{im}^{*}(\mathbf{r})\Delta
U(\mathbf{r})\psi_{jn}(\mathbf{r})
\end{align}
with $\Delta
U(\mathbf{r})=H_{SL}(\mathbf{r})-\sum_{\mathbf{R}_{i}}H_{QD}(\mathbf{r}-\mathbf{R}_{i})$, where
$H_{SL}$ is the superlattice Hamiltonian and $H_{QD}$ is the Hamiltonian of the isolated QD. 
The Hamiltonian \eqref{eq:ham} is used in the equations for the steady state non-equilibrium Green's functions for
the device states, which read
\begin{align}
\mathbf{G}^{R}(E)&=\Big[\big\{\mathbf{G}_{0}^{R}(E)\big\}^{-1}
-\boldsymbol{\Sigma}^{R I}(E)-
\boldsymbol{\Sigma}^{RC}(E)\Big]^{-1},\label{eq:retgf}\\
\mathbf{G}^{\lessgtr}(E)&=\mathbf{G}^{R}(E)
\big[\boldsymbol{\Sigma}^{\lessgtr I}(E)
+\boldsymbol{\Sigma}^{\lessgtr
C}(E)\big]\mathbf{G}^{A}(E),\label{eq:corrf}\\
\mathbf{G}_{0}^{R}(E)&=\left[(E+i\eta)\mathbbm{1}-\mathbf{H}_{0}\right]^{-1},\quad
\mathbf{G}^{A}(E)=[\mathbf{G}^{R}(E)]^{\dagger},
\end{align}
where $\mathbf{G}\equiv\left[G_{im,jn}\right]$ with
$G_{im,jn}(E)=\int d\tau e^{iE\tau/\hbar}G_{im,jn}(\tau),~\tau=t'-t,$
and $G_{im,jn}(t,t')=-\frac{i}{\hbar}\langle
\hat{T}_{\mathcal{C}}\{\hat{d}_{im}(t)\hat{d}_{jn}^{\dagger}(t')\}\rangle$ the quantum-statistical
non-equilibrium average defined on the Keldysh contour $\mathcal{C}$ \cite{keldysh:65}.
In the above equations, the self-energy terms $\Sigma^{\alpha I}$ ($\alpha\in\{\lessgtr,R,A$\}) describe the
renormalization of the single charge carrier Green's function due to the interactions relevant for
the photovoltaic device operation, i.e., coupling to photons, phonons and other charge carriers, and
are obtained within many-body perturbation theory on the level of the self-consistent first Born
approximation. As these self-energies are functionals of the carrier Green's functions, they need to
be computed self-consistently with the latter. The additional self-energy $\Sigma^{\alpha C}$
encodes the coupling of the QD states to the states of the contacts, which in this case are assumed as bulk electrodes.
Details on the general derivation and explicit form of the self-energies can be found in \cite{ae:jcel_review}.

Once the NEGF have been determined, they immediately provide any physical observable on the single
particle level. For instance, the density of states is obtained
from the NEGF and the QD eigenfunctions via
\begin{align}
\mathscr{D}(\mathbf{r},E)=&\sum_{i,j}\sum_{m,n}A_{im,jn}(E)\psi_{im}^{*}(\mathbf{r})\psi_{jn}(\mathbf{r}),
\end{align}
where $\mathbf{A}\equiv i(\mathbf{G}^{R}-\mathbf{G}^{A})$ is the charge carrier
spectral function. In terms of the tight-binding interdot hopping and Green's function
elements, the current flowing between two quantum dots can be
expressed as\footnote{The trace is over orbital indices.} 
\begin{align}
J_{i}=\frac{2e}{\hbar}\int
dE~\mathrm{tr}\left\{\mathbf{t}_{ii+1}\mathbf{G}^{<}_{i+1i}(E)-\mathbf{G}^{<}_{ii+1}(E)\mathbf{t}_{i+1i}\right\}.
\end{align}

\section{Implementation for Si-SiO$_{2}$-SiC QD heterostructure absorbers}
Since the wave functions and energies of the quantum dot eigenstates are determined separately for each
individual dot, the corresponding computational domain is small enough to allow, in principle, for the use of accurate 
ab-initio methods. In this paper, however, since the main focus is on the general formalism, the 
orbitals are approximated via the superposition of three separate 1D solutions for a simple one-band
effective mass model. In order to agree with full real space calculations, the local variation of the electrostatic
mean-field potential in principle needs to be included in the solution of the Schr\"odinger
equation, and to be computed self-consistently with the carrier density via Poisson's equation. In
this first approach, a linear potential drop resulting in a constant field is assumed and the orbitals are
computed only once for the flat-band case. This means that quantum-confined Stark effects resulting
from the distortion of the orbitals due to the local field are not considered at the present stage.
\begin{figure}[!t]
\begin{center}
\includegraphics[height=5.3cm]{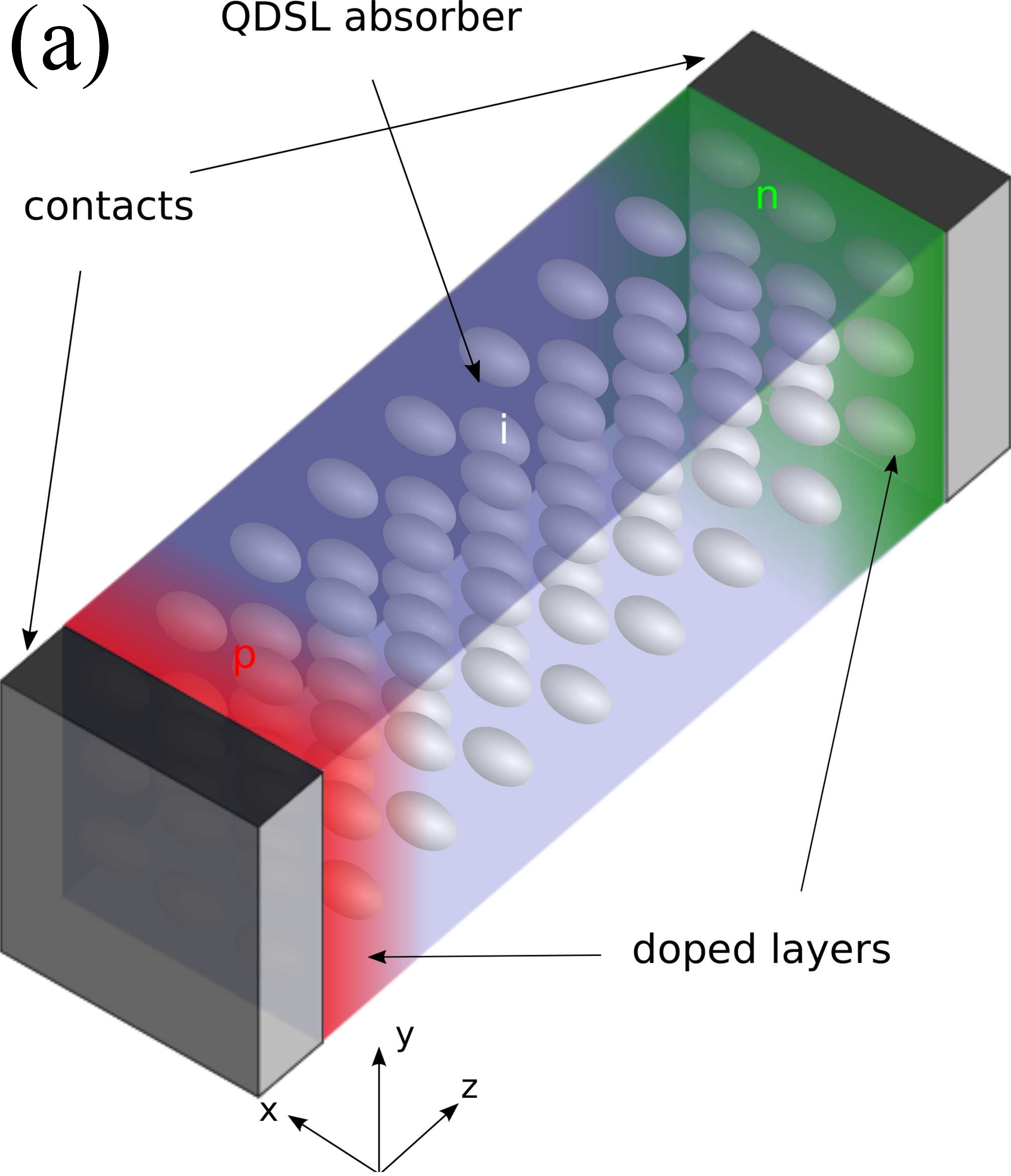}\quad\includegraphics[height=5.8cm]{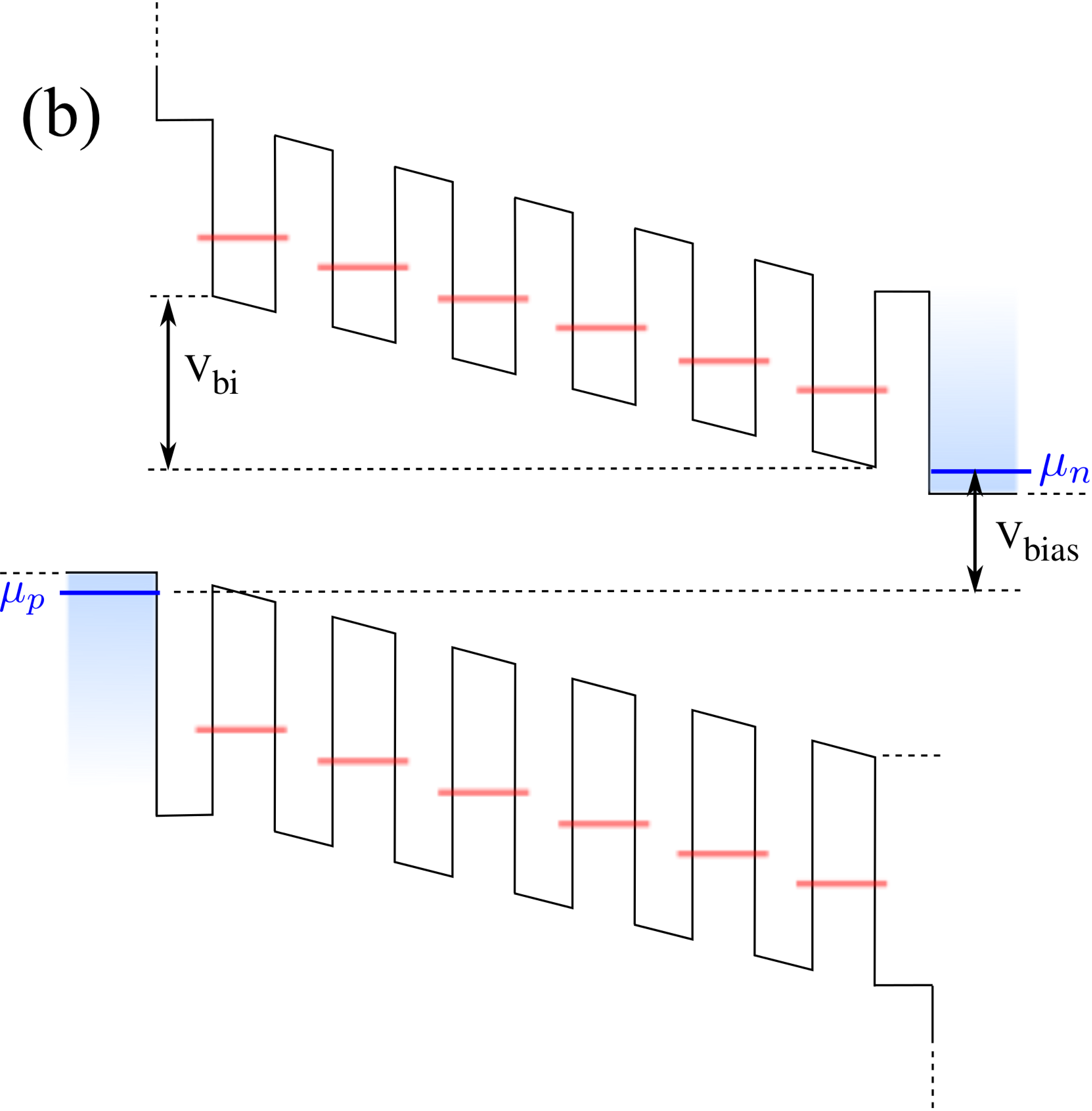}
\caption{\label{fig:struct_banddiag}(Color online)(a) Schematic structure and (b) resulting band
diagram of a $p$-$i$-$n$- QD-array absorber solar cell device. The resulting net potential drop
$V_{bi}$ in the absorbing region is determined by the combination of the built-in potential due to 
doping and the terminal bias voltage $V_{bias}$, the latter corresponding to the separation of the chemical potentials 
$\mu_{n,p}$ at the contacts.}
\end{center}
\end{figure}

The structure under investigation is shown schematically in Fig.~\ref{fig:struct_banddiag}(a). It 
consists of decoupled arrays of 20 cubic silicon quantum dots of 3 nm side length embedded laterally
in SiO$_{2}$ and vertically separated by thin SiC layers of 1~nm thickness\footnote{A similar structure was recently
fabricated as a candidate for an all-silicon tandem solar cell component \cite{ding:energy_procedia11}.}. The QD array
is coupled to contacts on both ends, and the regions adjacent to the contacts are doped to provide a built-in field. The
non-selfconsistent band 	diagram is shown in Fig.~\ref{fig:struct_banddiag}(b). The internal field depends on the
built-in potential $V_{bi}$, determined by the doping, and the bias voltage $V_{bias}$ applied at the right contact, the latter 	
corresponding to a finite separation of the chemical potentials associated with the equilibrated electrodes.
In order	to prevent leakage current flow and to enable charge separation even in absence of an internal electric field, minority carrier
contacts are closed artificially, the contacts thus becoming carrier-selective.

At this stage of implementation, the optical modes considered are those of a homogeneous medium and
with monochromatic occupation, and only radiative recombination is considered. Dissipation of energy is modelled via
coupling to a single optical (bulk) phonon mode described within the deformation potential approximation. The
energy of the optical branch was chosen at $E_{phon}=0.06$ eV. Additional broadening of the electronic states is obtained from elastic
scattering with	 acoustic phonons, using again the appropriate bulk deformation potentials.

\section{Numerical results}

\begin{figure}[!t]
\begin{center} 
\includegraphics[width=10cm]{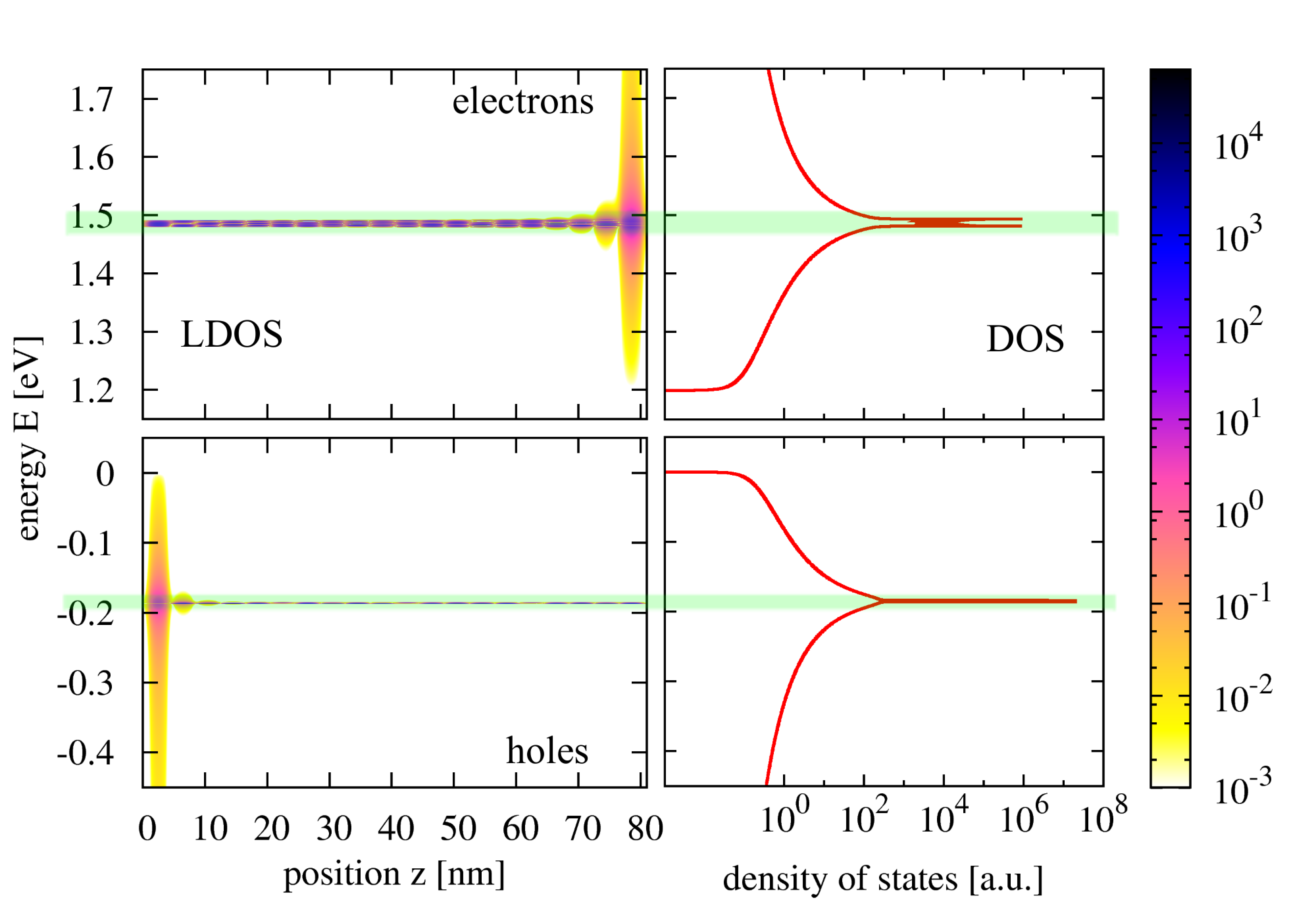}
\caption{\label{fig:ldos_intrinsic1}(Color online) Local density of states (LDOS) and integrated DOS 
of the selectively contacted QD-array structure from Fig.~\ref{fig:struct_banddiag} for vanishing internal
field ($V_{bi}=0$, flat ''miniband''). Only the contribution from the lowest orbital is shown. 
While the weak coupling of the QDs results in a very narrow band, the LDOS close to the carrier selective  contacts
shows a characteristic broadening due to hybridization with the states of the bulk electrode. The magnification of the 
shaded area is shown in Fig. \ref{fig:ldos_intrinsic2}}
\end{center}
\begin{center} 
\includegraphics[width=10cm]{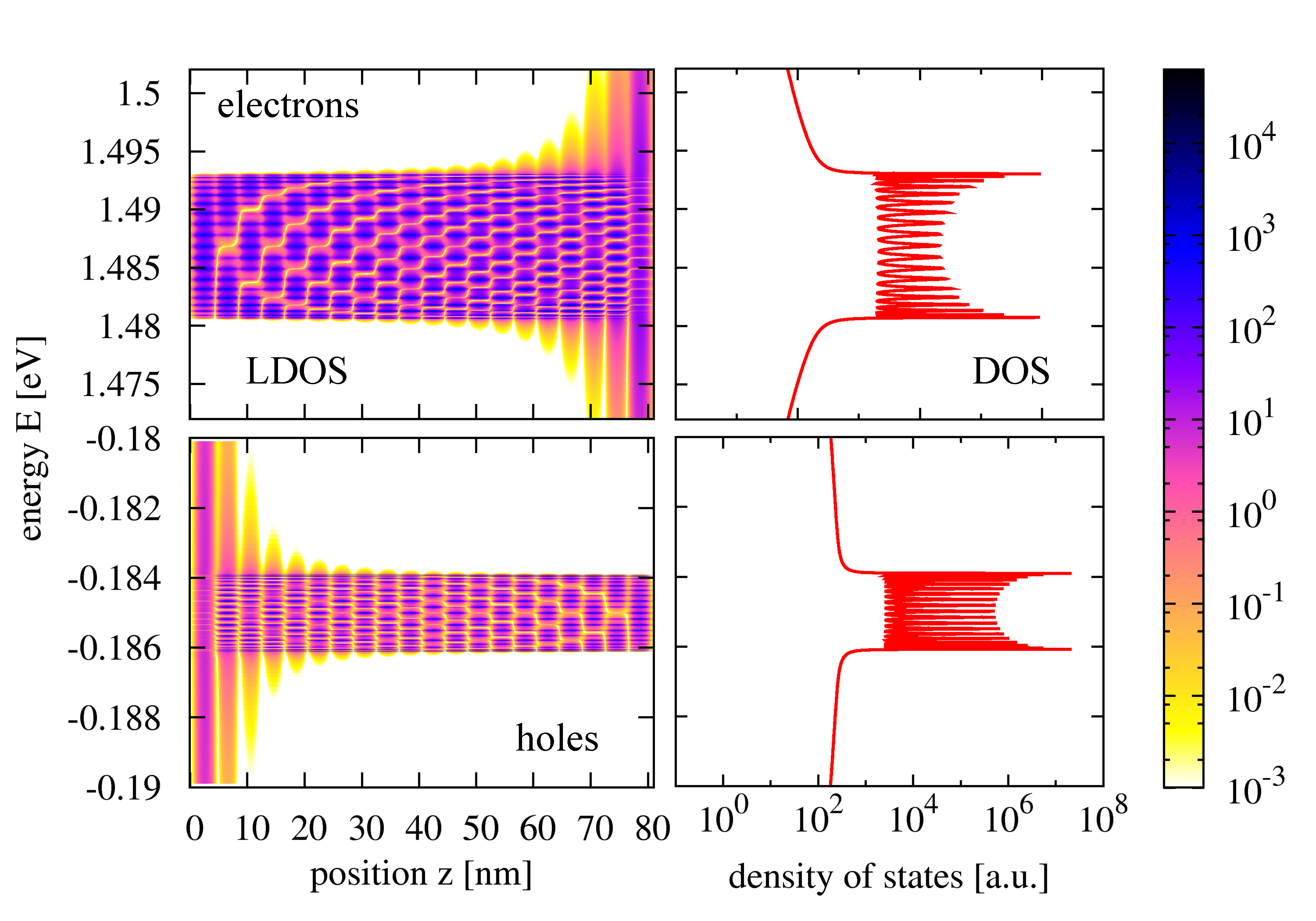}
\caption{\label{fig:ldos_intrinsic2} LDOS and integrated DOS at vanishing built-in field (shaded
area from Fig. \ref{fig:ldos_intrinsic1}). In this situation, the states are maximally delocalized and
extend over entire multi-QD structure. The discrete set of maxima in the LDOS on the QD positions
reflect both coupling and finite number of QDs in the structure.}
\end{center}
\end{figure}

Fig. \ref{fig:ldos_intrinsic1} shows the local as well as the spatially integrated density of states (LDOS and DOS)
of the selectively contacted QD-array structure from Fig.~\ref{fig:struct_banddiag} for vanishing internal
field ($V_{bi}=V_{bias}=0$), in the case where only the lowest QD orbital is considered. 
Due to the finite number of dots, the density of states shows discrete maxima, but
as a consequence of the flat band situation, the states are completely delocalized over the
structure, such that on every QD site, the multiplicity of states amounts to the number of QDs, as
can be 	recognized in the magnification of the shaded region (Fig. \ref{fig:ldos_intrinsic2}). The
difference 	between electrons and holes is due to the differing values for effective masses and
barrier height. While the weak mutual coupling of QD states results in a very narrow ''miniband'' in
the active device region, the LDOS is strongly broadened close to the carrier selective contacts,
which is due to the efficient coupling to the broad-band bulk electrode DOS. This hybridization with 
bulk electrode states is also reflected in the integrated DOS. 

\begin{figure}[!t]
\begin{center}
\includegraphics[width=10cm]{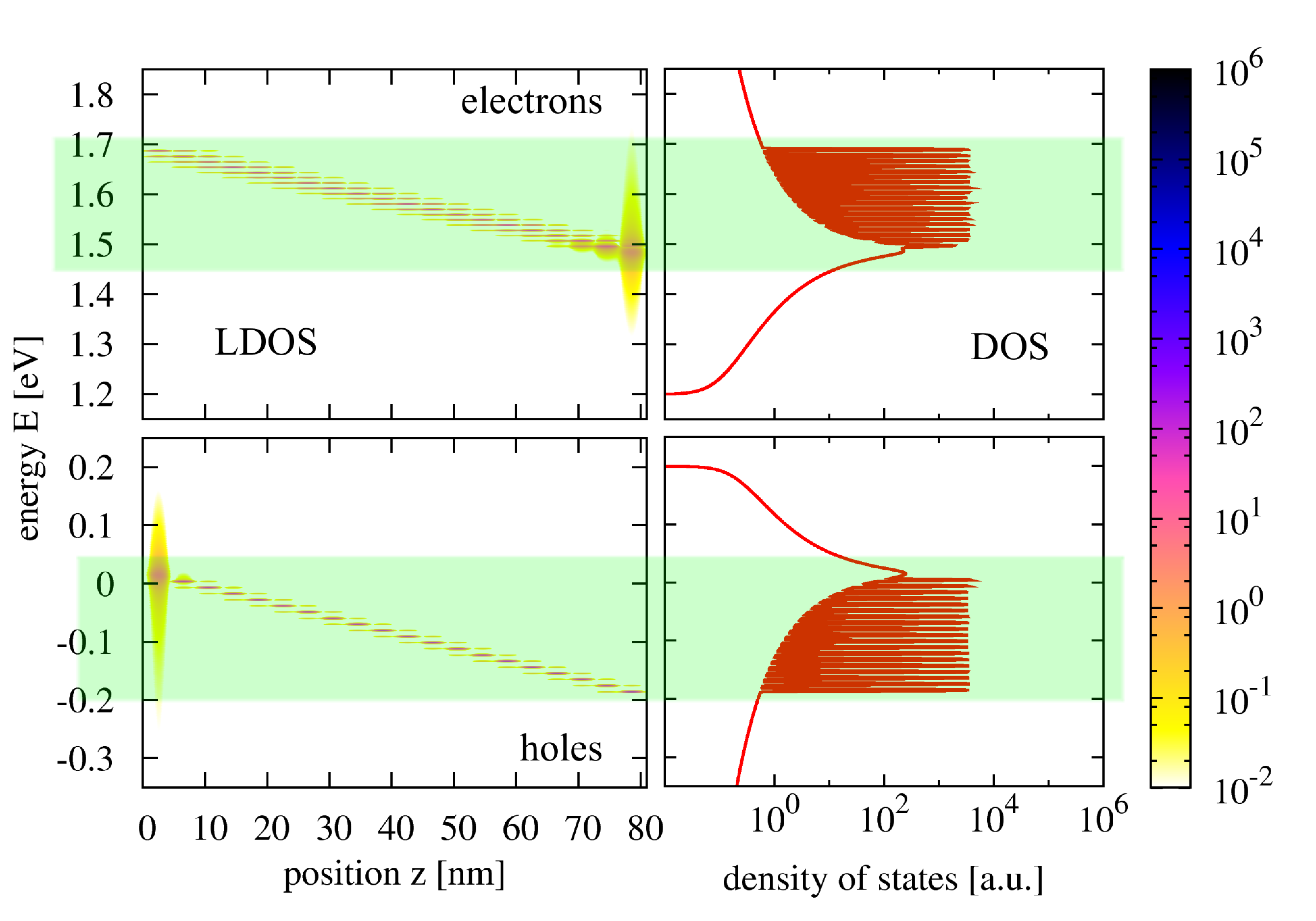}
\caption{\label{fig:ldos_field1}(Color online) Same as Fig.~\ref{fig:ldos_intrinsic1}, but for a
finite internal field of 25 kV/cm. The field splits the quasi-miniband into a broader level
structure with reduced local energetic overlap.}
\end{center}
\begin{center}
\includegraphics[width=10cm]{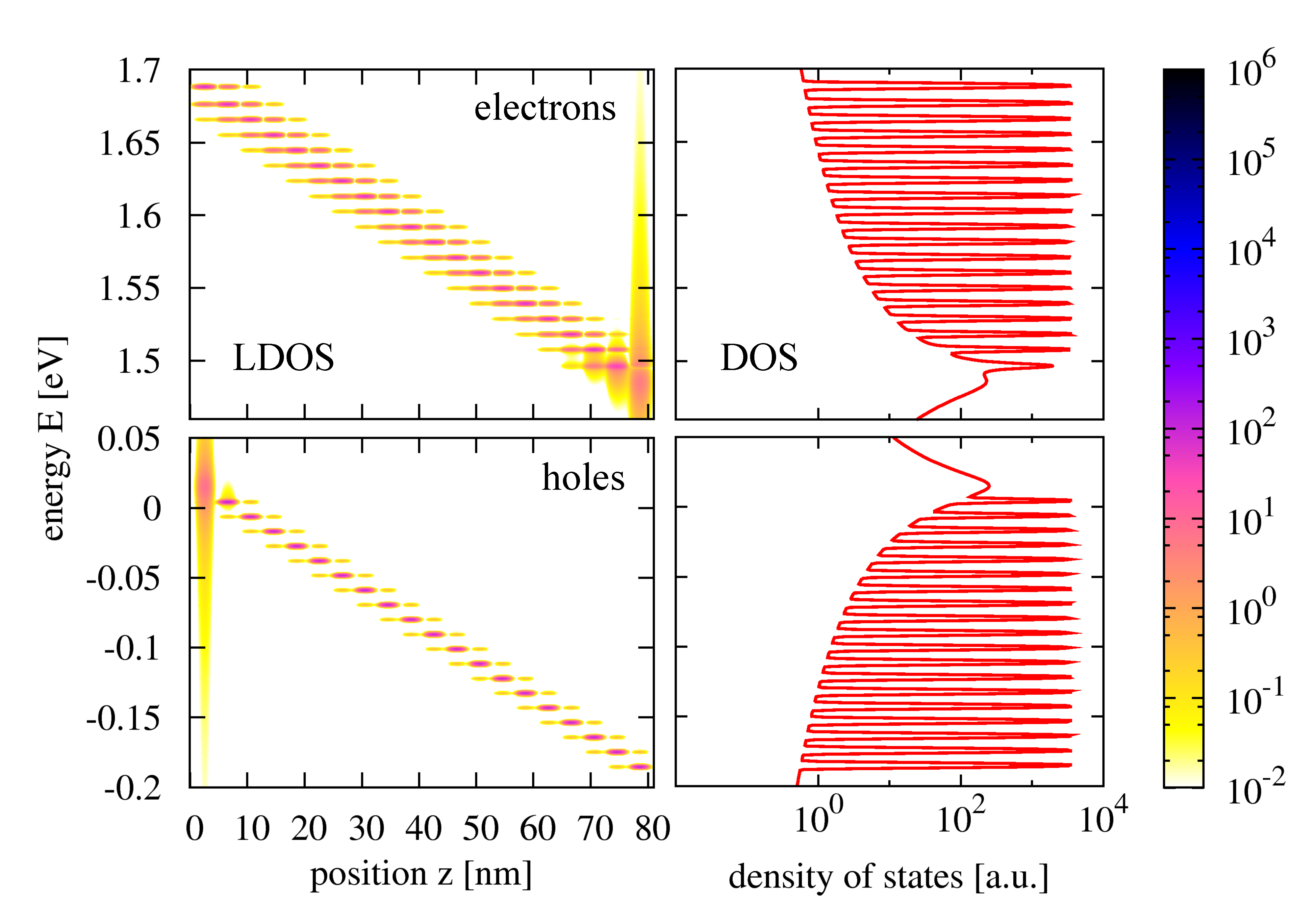}
\caption{\label{fig:ldos_field2}(Color online) Detailed view of the situation at
finite internal field of 25 kV/cm (shaded area in Fig. \ref{fig:ldos_field2}). The field has a strong
localizing effect on the QD-array states, which are spatially confined to a small number of
neighboring QD. Without the presence of higher states, ballistic transport to the contact is no
longer possible in this situation, i.e., energy dissipation via inelastic scattering processes is
required to enable photocarrier extraction.}
\end{center}
\end{figure}

In the flat band situation, extraction of photogenerated charge carriers is efficient, since
all the absorbing states are directly connected to the contact states. However, the situation changes
drastically in the presence of a finite field, as displayed in Fig.~\ref{fig:ldos_field1}. The
width of the ''miniband'' increases in the integrated DOS, but in the LDOS, the number of states per
QD with significant spectral weight is reduced, since even for a field as weak as 25 kV/cm, which is
close to the values encountered in real devices, the states are spatially localized on a few QDs.
This carrier localization, clearly seen in the magnification shown in Fig.~\ref{fig:ldos_field2},
prevents a direct ballistic extraction of photogenerated charge carriers and thus increases the
significance of scattering assisted transport ranging from sequential tunneling to hopping between
single dots.

\begin{figure}[t]
\begin{center}
\includegraphics[width=12cm]{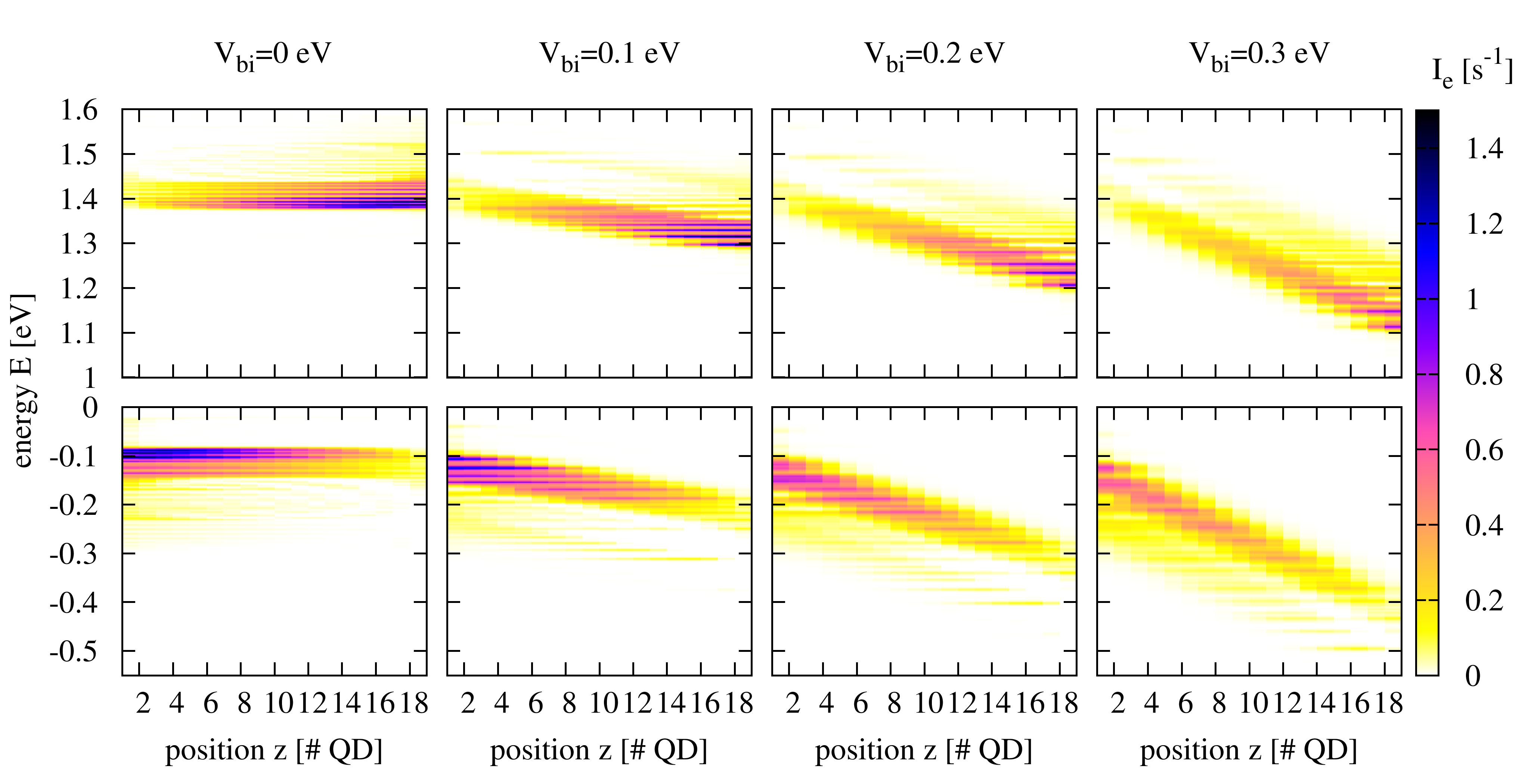}
\caption{\label{fig:photocurrent}(Color online) Photocurrent spectrum at photon energy of 1.67 eV
for zero applied bias and different values of the built-in field. Due to strong electron-phonon
interaction, the photogenerated electron-hole pairs relax to the effective band edge within the
delocalization range of a few quantum dot spacings, and photocarrier extraction remains possible
even in absence of ballistic current contributions. In spite of the large variation in the
current spectrum for different internal fields, the energy integrated current differs only slightly,
and the total current (electron+hole) is perfectly conserved over the entire device.}
\end{center}
\end{figure}

The restriction to the lowest orbital leads to an underestimation of the overall inter-dot
coupling. In order to study the extraction of excess charge generated under illumination,
the single-band hopping matrix element is increased to 0.05 eV. To investigate the transport of photogenerated
charge carriers under conditions were the states are no longer completely delocalized, i.e., in the
situation where inelastic scattering is crucial to enable carrier flow between quantum dots, the
local photocurrent spectrum is computed for gradually increasing internal field. The result is displayed 
in Fig.~\ref{fig:photocurrent} for monochromatic illumination at a photon energy of 1.67 eV,
corresponding roughly to the effective energy gap. Even though the current spectrum is strongly
modified by the internal field, the energy integrated current differs only slightly for different field values, and the
overall current, i.e. the sum of electron and hole contributions, is perfectly conserved over the entire device. The
strong electron-phonon scattering leads to a fast relaxation within the few QDs over which the states are
delocalized, such that photocurrent flow follows the effective band edge. Thus, efficient
photocarrier extraction is recovered at the expense of the energy dissipated in the inelastic
scattering process.

\section{Conclusions}
The QDTB-NEGF approach extends the applicability of powerful quantum-kinetic methods to 
optoelectronic devices based on electronically coupled quantum dot structures, with arbitrary
internal fields and spatial arrangement. Application to an array of weakly coupled quantum dot
reveals the significant carrier localization for finite electric fields in the absorbing region. For
stronger coupling, photogenerated carriers are extracted efficiently also at considerable internal
fields via phonon-assisted sequential tunneling and inter-dot hopping.

\begin{acknowledgements}
Financial support was provided by the German Federal Ministry
of Education and Research (BMBF) under Grant No. 03SF0352E.
\end{acknowledgements}

\bibliographystyle{spphys}       

\bibliography{manuscript}

\end{document}